\documentclass[prb,twocolumn]{revtex4-1}
\usepackage{amsmath,graphicx}

\begin{document}
\title{Chiral Majorana fermion modes regulated by a scanning tunneling
    microscope tip}
\author{Yan-Feng Zhou$^{1,2}$}
\author{Zhe Hou$^{1,2}$}
\author{Ying-Tao Zhang$^{3}$}
\author{Qing-Feng Sun$^{1,2,4}$}\email{sunqf@pku.edu.cn}

\affiliation{$^{1}$International Center for Quantum Materials, School of Physics, Peking University, Beijing 100871, China}
\affiliation{$^{2}$Collaborative Innovation Center of Quantum Matter, Beijing 100871, China}
\affiliation{$^{3}$College of Physics, Hebei Normal University, Shijiazhuang 050016, China}
\affiliation{$^{4}$CAS Center for Excellence in Topological Quantum Computation, University of Chinese Academy of Sciences, Beijing 100190, China}

\date{\today}
\begin{abstract}
The Majorana fermion can be described by a real wave function with only two phases (0 and $\pi$) which provide a controllable
degree of freedom. We propose a strategy to regulate the phase of the chiral Majorana state by coupling with
a scanning tunneling microscope tip in a system consisting of quantum anomalous Hall insulator coupled with a superconductor.
With the change of the chemical potential, the chiral Majorana state
can be tuned alternately between 0 and $\pi$, in which, correspondingly, the perfect normal tunneling and
perfect crossed Andreev reflection appear.
The perfect crossed Andreev reflection, by which a Cooper pair can be split into two electrons
going into different terminals completely,
leads to a pumping current and distinct quantized resistances.
These findings may provide a signature of Majorana fermions and pave a feasible avenue to regulate the phase of Majorana state.
\end{abstract}

\maketitle
\section{\label{sec1}Introduction}

Topological superconductors (TSCs), the superconducting counterparts of topological insulators,
have attracted more and more attentions for catching Majorana fermions in condensed matter systems\cite{ReadN,QiXL1,Alicea,Beenakker,ElliottSR}.
The Majorana zero modes exhibiting non-Abelian statistics\cite{ReadN,IvanovDA,NayakC}
can exist in the core of superconducting vortices in the chiral TSCs and have potential
applications in topological quantum computation. Exotic effects, such as the $4\pi$-periodic
Josephson effect\cite{FuL2,Lutchyn}, nonlocal tunneling\cite{Nilsson,Law,WangZ},
can be the promising manifestation of the zero-dimensional Majorana bound states.
Theoretical proposals have shown that the chiral TSCs can be realized by inducing superconductivity
on quantum Hall systems, quantum anomalous Hall insulators (QAHI)\cite{QiXL2,WangJ},
and two-dimensional systems with spin-orbit coupling\cite{FuL1,SauJD}
via the proximity effect of an s-wave superconductor. Moreover, there also exist $\mathcal{N}$
one-dimensional chiral Majorana edge modes (CMEMs) in the interface of chiral TSCs with  topological
Chern number $\mathcal{N}$. Especially, a half-integer conductance plateau at the coercive field
in a hybrid TSC/QAHI structure based on the one-dimensional CMEMs has been predicted by theoretical studies\cite{WangJ,Chung,Lian,nadd6}
and been observed recently in a transport experiment\cite{HeQL}, providing a transport signature of CMEM.\cite{WangJ,nadd7}
For promising applications of Majorana fermions, it is important to control the Majorana modes
in a feasible way and explore more compelling experimental evidences of them in the chiral TSCs\cite{GMenard,Tiwari}.

The electron-hole conversion can occur at the interface between a superconductor and a conductor,
forming a Cooper pair into the superconductor. Here the incoming electron can be reflected as
a hole in the same lead, known as local Andreev reflection (LAR)\cite{Andreev,asun1}
or be scattered to the other lead, known as crossed Andreev reflection
(CAR)\cite{Byers,Deutscher,nadd2,Recher}.
When bias below the superconductor gap, the transport properties of the system
are mainly determined by the Andreev reflections.\cite{asun1,nadd1}
Because of the LAR and other processes, such as normal reflection and normal tunneling,
the probability of the CAR is in general very small. Recently, some works have
focused on the Andreev reflections in the superconductor/togological system and some exotic phenomena have been
predicted.\cite{Chen,Veldhorst,Cayssol, James, ZhangYT, WangJ2,addDLoss,addPachos,nadd3,nadd4,nadd5}
For example, a resonant CAR can be obtained with other processes
being prohibited, through band engineering of electron/hole in the leads\cite{Chen,Veldhorst,Cayssol}
or with the assistance of the Majorana end sates\cite{James}.
Besides, by utilizing the unidirectionality of the
topologically-protected edge states, the quantized CAR is proposed in the systems
by coupling an s-wave superconductor with QAHI ribbon\cite{ZhangYT} and spin-valley topological insulator\cite{WangJ2}.

In this paper, we propose an avenue to control the phase of the CMEMs in the hybrid TSC/QAHI system by using a
scanning tunneling microscope (STM) tip and demonstrate a quantized perfect CAR caused by the phase-regulating.
Because of the property that the Majorana fermion is a self-Hermitian particle, its wave function is real and
its phase can only be $0$ or $\pi$. Moreover, for one-dimensional chiral Majorana fermion with propagation
velocity $v_M$ obeying the Hamiltonian $H = -i\hbar v _M\partial_x$,\cite{FuL3,Akhmerov,Park} the current
density $j_x = v_M\Psi^*\Psi$ and the wave function $\Psi$ is nonzero at any site due to the current
conservation condition. Consequently, the phase of Majorana fermion $\gamma$ propagating forward along
the CMEM is zero only, i.e. $\gamma \rightarrow \gamma$.\cite{FuL3,Akhmerov} However, with the branch cut
introduced by the STM tip, the chiral Majorana fermion cannot be regarded as one-dimensional system
any more, leading that its phase can be 0 or $\pi$. We show that the phase can be easily regulated from $0$ to $\pi$,
i.e. $\gamma \rightarrow -\gamma$ for phase $\pi$. Corresponding to the phases $0$ and $\pi$,
the perfect normal tunneling and quantized perfect CAR can occur, respectively.

The organization of this paper is as follows. After this introductory section, Sec.~\ref{sec2}
describes the theoretical models of the TSC/QAH system coupled with STM tip and the methods
for calculating the tunneling coefficient, the LAR coefficient, the CAR coefficient, and the
current. Sec.~\ref{sec3} presents the numerical results on the details of phase modulation
of CMEMs by STM tip, coherence, and experimental signatures. Sec.~\ref{sec4} concludes this paper.
Some auxiliary materials are relegated to Appendix.

\section{\label{sec2}Model}

We consider a hybrid TSC/QAHI system where a chiral TSC
island is introduced near one edge of the QAHI ribbon and a STM tip couples to
the TSC island, as shown in Fig.1 (a). In fact, some recent experimental and theoretical
works have applied the STM tip to probe the Majorana fermion.\cite{GMenard,nadd8,nadd9}
Here we apply it to regulate the phase of the Majorana state propagating forward
along the CMEM.
The QAHI states have been predicted in some realistic proposals by doping
topological insulators with magnetic dopants and experimentally realized in
Cr-doped\cite{ChangCZ1,Checkelsky,Kou,Bestwick,Kandala} or
V-doped\cite{ChangCZ2} $\mathrm{(Bi,Sb)_2Te_3}$ magnetic topological insulator thin films.
For the low-energy states near the $\Gamma$ point, the two-band Hamiltonian describing the QAHI state
can be expressed as\cite{QiXL2} $\mathcal{H}_{\mathrm{QAHI}}=\sum_\mathbf{p}\psi_\mathbf{p}^\dagger H_{\mathrm{QAHI}}(\mathbf{p})\psi_\mathbf{p}$,
with $\psi_\mathbf{p}=(c_{\mathbf{p}\uparrow},c_{\mathbf{p}\downarrow})^T$ and,
 \begin{equation}\label{HQ}
 H_{\mathrm{QAHI}}(\mathbf{p})= (m+Bp^2)\sigma_z+A(p_x\sigma_x+p_y\sigma_y),
 \end{equation}
where $c_{\mathbf{p}\sigma}$ and $c_{\mathbf{p}\sigma}^\dag$ are, respectively, the annihilation
and creation operators with momentum $\mathbf{p}$ and spin $\sigma=\uparrow,\downarrow$, and $\sigma_{x,y,z}$ are
Pauli matrices for spin. $A$, $B$ and $m$ are material parameters. For the numerical calculation,
the Hamiltonian $\mathcal{H}_{\mathrm{QAHI}}$ can be further mapped into a square lattice model
in the tight-binding representation\cite{adatta},
\begin{align}\label{HQL}
\mathcal{H}_{\mathrm{QAHI}} = \sum_\mathbf{i}[\psi_\mathbf{i}^{\dag} T_0 \psi_\mathbf{i}+(\psi_\mathbf{i}^{\dag}T_x\psi_{\mathbf{i}+\delta \mathbf{x}}+\psi_\mathbf{i}^{\dag} T_y \psi_{\mathbf{i}+\delta \mathbf{y}})+\mathrm{H.c.}],
\end{align}
with
$T_0 =(m+4B\hbar^2/a^2)\sigma_z$,
$T_x = -(B\hbar^2/a^2)\sigma_z-(iA\hbar/2a)\sigma_{x}$ and
$T_y =-(B\hbar^2/a^2)\sigma_z-(iA\hbar/2a)\sigma_{y}$.
Here $\psi_\mathbf{i}=(c_{\mathbf{i}\uparrow},c_{\mathbf{i}\downarrow})^T$,
$c_{\mathbf{i}\sigma}$ and $c_{\mathbf{i}\sigma}^\dag$ are, respectively,
the annihilation and creation operators on site $\mathbf{i}$ with spin $\sigma$.
$a$ is the lattice length and $\delta \mathbf{x}$ ($\delta \mathbf{y}$) is the unit vector along $x$ ($y$) direction.
The topological properties of the Hamiltonian $\mathcal{H}_{\mathrm{\mathrm{QAHI}}}$ are determined by the sign of $m/B$.
For $m/B<0$, the QAHI state with Chern number $\mathcal{C}=1$ is obtained and there exists one chiral edge mode at each edge in a QAHI ribbon
as indicated by black arrows in Fig.1 (a). For $m/B>0$, the Hamiltonian $\mathcal{H}_{\mathrm{QAHI}}$ describes a normal insulator state
with Chern number $\mathcal{C}=0$.

\begin{figure}
\centering
\includegraphics[width=.9\linewidth]{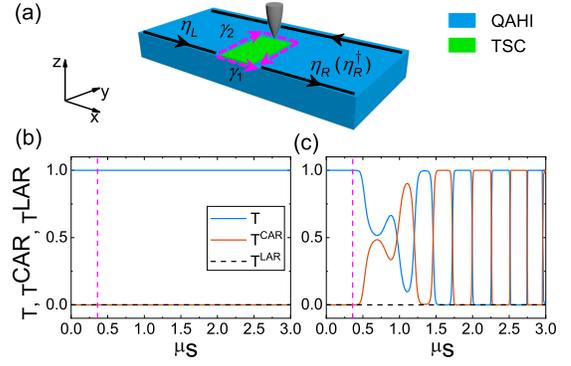}
\caption{(color online) (a) Schematic diagram of the hybrid TSC/QAHI system coupled by a STM tip.
In the green region, a TSC state is induced through the proximity effect by coupling with an s-wave superconductor. The grey tip is the STM tip.
Black arrows label the QAHI edge states and magenta arrows indicate the CMEMs.
(b) and (c) show the normal tunneling coefficient $T$, CAR coefficient $T^{CAR}$, and LAR coefficient $T^{LAR}$ as functions of $\mu_s$
without and with the STM tip, respectively.
The STM tip coupling strength $\Gamma=1$, the coupling position is at the middle of the upper edge of
the TSC island, and the contact region is $3\times3$ lattices in the numerical simulation.
The QAHI ribbon width $W = 150a$ and the size of the TSC island is
$(L_x, L_y)= (20a, 90a)$.
}
\label{fig:fig1}
\end{figure}

Near one edge of the QAHI ribbon, an s-wave superconductor is coupled to it
[see the green region in Fig.1 (a)]
and the proximity effect can induce a finite pairing potential $\Delta$
in the superconductor-covered QAHI region. In this region, the electron and hole excitations
are described by the Bogoliubov-de Gennes (BdG) Hamiltonian,
$\mathcal{H}_{\mathrm{BdG}}=\frac{1}{2}\sum_\mathbf{p}\Psi_\mathbf{p}^\dagger H_{\mathrm{BdG}}(\mathbf{p})\Psi_\mathbf{p}$,
in the basis of $\Psi_\mathbf{p}=(c_{\mathbf{p}\uparrow},c_{\mathbf{p}\downarrow},c_{\mathbf{-p}\uparrow}^{\dag},c_{\mathbf{-p}\downarrow}^{\dag})^T$, and
\begin{equation}\label{HBDG}
H_{\mathrm{BdG}}=\left(
                      \begin{array}{cc}
                        H_{\mathrm{QAHI}}(\mathbf{p})-\mu_s & i\Delta\sigma_y \\
                        -i\Delta^{\ast}\sigma_y & -H_{\mathrm{QAHI}}^{\ast}(-\mathbf{p})+\mu_s \\
                      \end{array}
                    \right)
.
\end{equation}
Here a finite chemical potential $\mu_s$ has been taken into account inside the TSC island.
Using Eq.(\ref{HQL}), the lattice version of Hamiltonian $\mathcal{H}_{\mathrm{BdG}}$ can also be obtained.
According to the Altland-Zirnbauer symmetry classification scheme, the Hamiltonian
$\mathcal{H}_{\mathrm{BdG}}$ which has intrinsic particle-hole symmetry,
but no time reversal symmetry belongs to the D class TSC\cite{Schnyder}.
The D class TSCs in two-dimension can be described by Chern number $\mathcal{N}$ and
support $\mathcal{N}$ CMEMs. According to the phase diagram of Hamiltonian
$\mathcal{H}_{\mathrm{BdG}}$,\cite{QiXL2} for a finite superconductor gap $\Delta$ and negative $m$,
the TSC region undergoes a phase transition from $\mathcal{N}=2$ to $\mathcal{N}=1$
(also called chiral TSC) by changing chemical potential $\mu_s$ and the phase boundary is determined
by the condition: $\Delta^2+\mu_s^2=m^2$. It is worth noting that here $\mathcal{N}=2$
CMEMs are topologically equivalent to one QAHI edge state ($\mathcal{C}=1$).
Very recently, the TSC with $\mathcal{N}=1$ has been successfully realized in the experiment\cite{HeQL}.

Then the Hamiltonian $H$ of the whole setup consisting of the hybrid TSC/QAHI ribbon coupled
by a STM tip [see Fig.1 (a)] is
\begin{eqnarray}
 H= \mathcal{H}_{\mathrm{QAHI/BdG}} + H_{\mathrm{STM}} +H_\mathrm{C},
\end{eqnarray}
where $\mathcal{H}_{\mathrm{QAHI/BdG}}$, $H_{\mathrm{STM}}$ and $H_\mathrm{C}$ are the Hamiltonian of the hybrid TSC/QAHI ribbon, the STM tip and the coupling between them, respectively.
$\mathcal{H}_{\mathrm{QAHI/BdG}}$ has been shown in Eqs.(1-3).
The Hamiltonians $H_{\mathrm{STM}}$ and $H_\mathrm{C}$ are
\begin{equation}\label{STM}
H_{\mathrm{STM}} +H_{\mathrm{C}}=
\sum_{\mathbf{i},k}(\varepsilon_{\mathbf{i}k}a_{\mathbf{i}k}^{\dag}a_{\mathbf{i}k}+\\
                 t_da_{\mathbf{i} k}^{\dag}\psi_{\mathbf{i}}+\mathrm{H.c.}),
\end{equation}
in which, $a_{\mathbf{i}k}=(a_{\mathbf{i}\uparrow k},a_{\mathbf{i}\downarrow k})^T$,
$a_{\mathbf{i}\sigma k}$ and $a_{\mathbf{i}\sigma k}^\dag$ are, respectively,
the annihilation and creation operators of the STM tip with spin $\sigma$.
Here the STM tip couples to several sites $\mathbf{i}$ only
and $t_d$ is the coupling strength. The coupling strength
is also characterized by $\Gamma=2\pi\rho t_d^2$ with $\rho$ being
the density of states of the STM tip.

Note that, here the Hamiltonian of the STM tip is the same as that of a metallic lead.
So the effect of the STM tip can be produced by coupling a metallic island to
the TSC island in the experiment and the similar results can be also obtained.\cite{Tiwari}
But in a specific setup, the coupling strength between the metallic island
and the TSC island cannot be changed.
By contrast, the STM tip is moveable, and the coupling strength $\Gamma$
between the tip and the TSC island can be controllable.

In the numerical calculation, we set $m=-0.5$, $\Delta=0.35$, $A=B=1$ with
a regularization lattice constant $a=1$ and $\hbar=1$. For an estimation in real materials,
$\hbar\nu_F\sim260\ \mathrm{meV\ nm}$ ($\nu_F$ is Fermi velocity) and the proximity effect
induced superconductor gap $\Delta=0.35\ \mathrm{meV}$.\cite{WangJ}
The lattice constant $a=\hbar\nu_F/\mathrm{A}=0.26 \ \mathrm{\mu m}$ [See Eq.(\ref{HQL})]
and the TSC island size $(L_x, L_y) = (20a, 90a)=(5.2 \ \mathrm{\mu m}, 23.4\ \mathrm{\mu m})$ in Fig.1(a), where $L_x$ and $L_y$ are the length and width of the TSC island.
This size is similar with the experiment device\cite{HeQL}.
The normal tunneling coefficient $T$, CAR coefficient $T^{CAR}$,
and LAR coefficient $T^{LAR}$ can be calculated by using
the nonequilibrium Green's function method combined with Landauer-B\"{u}ttiker formula
(see Appendix A for details of the calculation\cite{asun1,adatta,asun2,asun3,alee}).

\section{\label{sec3}Results}
\subsection{Phase modulation by STM tip}

When without the STM tip, normal tunneling coefficient $T$, CAR coefficient $T^{CAR}$,
and LAR coefficient $T^{LAR}$ at the zero incident energy case
as functions of the chemical potential $\mu_s$ are shown in Fig.1 (b).
Here the normal tunneling is perfect with $T=1$,
but $T^{CAR}$, $T^{LAR}$, and normal reflection $R$ disappear regardless of
the chemical potential $\mu_s$.
For $\mu_s<\sqrt{m^2-\Delta^2}$ [the left region of the magenta dashed in Fig.1 (b)],
the TSC island is in the $\mathcal{N}=2$ phase which is topologically equivalent to the QAHI state.
In this case, the incident electron from the left side can be transmitted transparently
to the right side.
On the other hand, for $\mu_s >\sqrt{m^2-\Delta^2}$, the TSC island is in $\mathcal{N}=1$ phase
with a single CMEM. Now when the incident electron $\eta_L$ along the QAHI edge state
from the left side
arrives at the interface of the TSC and QAHI, it is separated into two Majorana fermions $\gamma_1$ and $\gamma_2$,
i.e. $\eta_L = \gamma_1 +i\gamma_2$ with $\gamma_1 =\frac{1}{2}(\eta_L +\eta^{\dagger}_L)$
and $\gamma_2 =\frac{1}{2i}(\eta_L -\eta^{\dagger}_L)$.
Then $\gamma_1$ and $\gamma_2$ propagate
forward along the lower and upper CMEMs, respectively, indicated by magenta arrows in Fig.1 (a).
Notice that the TSC island does not touch with the upper edge of the QAHI ribbon [Fig.1 (a)],
so $\gamma_1$ and $\gamma_2$ must meet at lower right corner of the TSC island,
and they combine into an ordinary fermion again to the right QAHI terminal.
Due to the reality constraint on the wave function of the one-dimensional CMEM,
when the Majorana fermion propagates forward along the CMEM, it cannot change its sign,
i.e. its phase can only be zero with $\gamma_1 \rightarrow \gamma_1$ and
$\gamma_2 \rightarrow \gamma_2$.\cite{FuL3,Akhmerov}
This means that the outgoing particle is $\gamma_1 +i\gamma_2 =\eta_R$, which is an electron.
Therefore the normal tunneling coefficient $T=1$ and
all other processes also disappear in the chiral TSC phase.

\begin{figure*}[!ht]
\centering
\includegraphics[width=15cm]{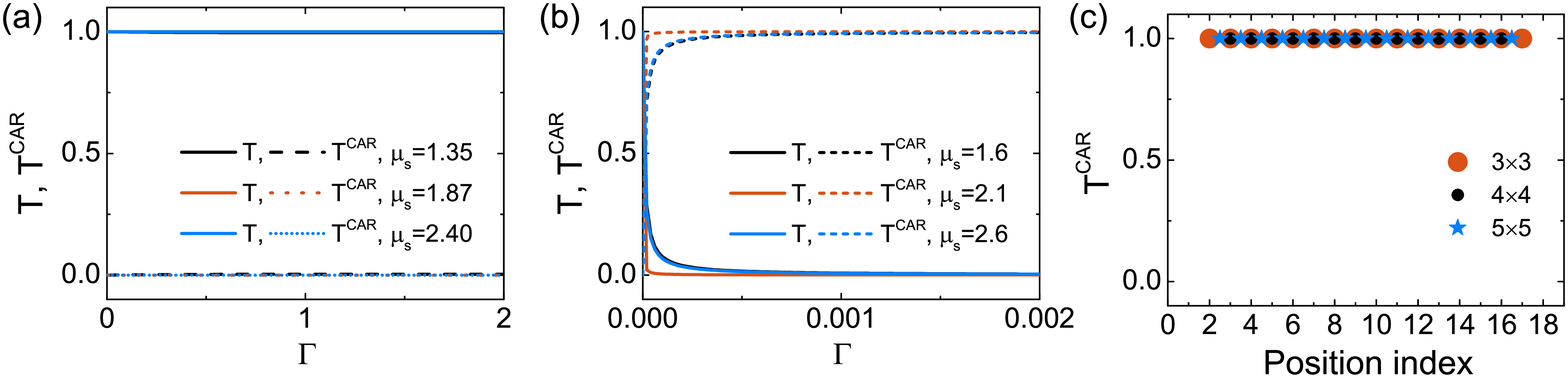}
\caption{(color online) (a) and (b) show the normal tunneling coefficient $T$ and
CAR coefficient $T^{CAR}$ versus the coupling strength $\Gamma$ of the STM tip for
several chemical potential $\mu_s$.
These $\mu_s$ in (a) and (b) are respectively the centers of the $T=1$ plateaus
and the $T^{CAR}=1$ plateaus in Fig.1 (c). In (a) three solid (dot) curves overlap together.
(c) is $T^{CAR}$ versus the coupling position of the STM tip with the different tip sizes
and $\mu_s=2.1$.
Here the position index from 0 to 19 means from left to right along
the upper edge of the TSC island [see Fig.1 (a)].
The other unmentioned parameters are the same as Fig.1 (c).
}
\label{fig:fig2}
\end{figure*}

In order to regulate the phase of Majorana state,
an STM tip is coupled to one edge of the TSC island [see Fig.1 (a)],
which can break the one-dimensional channel behavior due to the branch.
Now the phase can be $0$ or $\pi$, and the outgoing upper Majorana state can be $\gamma_2$ or
$-\gamma_2$ which depends on the chemical potential $\mu_s$.
When $\gamma_2 \rightarrow \gamma_2$, the outgoing particle $\eta_R = \gamma_1 +i\gamma_2 $ is an electron,
then the quantized perfect normal tunneling occurs with $T=1$ and $T^{CAR}=T^{LAR}=R=0$.
On the other hand, while $\gamma_2 \rightarrow -\gamma_2$, the outgoing particle is a hole:
$\gamma_1 - i\gamma_2 =\eta_R^{\dagger}$. The occurrence of the perfect electron-hole conversion
leads to the quantized perfect CAR effect. In this situation, the CAR coefficient $T^{CAR}$ is one and all other
processes (including the normal tunneling, normal reflection and LAR) are completely suppressed.
Fig.1 (c) shows the normal tunneling coefficient $T$, CAR coefficient $T^{CAR}$ and LAR coefficient $T^{LAR}$ versus chemical potential $\mu_s$ with the coupling of the STM tip.
As expected, with the increase of $\mu_s$, $T$ and $T^{CAR}$ appear alternately as plateaus
with the plateau values being one in the chiral TSC regime. Here both normal reflection and
LAR completely disappear because the TSC island does not touch the upper edge of QAHI ribbon,
so $T+T^{CAR}=1$. When $T^{CAR}=0$, the normal tunneling coefficient $T=1$,
it corresponds to the perfect tunneling. Whereas while
$T=0$, the CAR coefficient $T^{CAR}=1$, it is the quantized perfect CAR.
That is to say, under the coupling of the STM tip
the phase of the Majorana state can be adjusted to $0$ or $\pi$ by
tuning chemical potential $\mu_s$, which provides a different way to
introduce the branch cut from the interferometry of Majorana fermions by superconducting vortex\cite{FuL3,Akhmerov}.

Next we study the effect of the coupling strength $\Gamma$ of the STM tip on
$T$ and $T^{CAR}$.
Here we consider the current at the STM tip terminal being zero, i.e., the STM tip terminal is open.
Fig.2 (a) shows the normal tunneling coefficient $T$ and CAR coefficient $T^{CAR}$
versus coupling strength $\Gamma$ under several chemical potential $\mu_s$
which are at centers of the $T=1$ plateaus [see Fig.1 (c)].
$T$ maintains unity and $T^{CAR}$ is always zero with increasing $\Gamma$ from $0$,
indicating that at these $\mu_s$ the phase of the Majorana state is $0$ and
the perfect tunneling occurs regardless of the coupling of STM tip.
More interestingly, for other chemical potential $\mu_s$ which locate at
centers of the $T^{CAR}=1$ plateaus,
the normal tunneling coefficient $T$ and CAR coefficient $T^{CAR}$ are strongly
affected by the coupling of the STM tip [see Fig.2 (b)].
While the coupling strength $\Gamma=0$, $T=1$ and $T^{CAR}=0$ which corresponds to the perfect tunneling.
With the increase of $\Gamma$, the normal tunneling coefficient $T$ reduces from $1$ to $0$ and
the CAR coefficient $T^{CAR}$ increases from $0$ to $1$ rapidly.
$T$ reaches $0$ and $T^{CAR}$ is $1$ even for a very small $\Gamma$ (e.g. $0.001$),
which is the quantized perfect CAR effect. This means that by the coupling of the STM tip,
the phase of the Majorana state propagating along the CMEM can be regulated from $0$ to $\pi$,
and the perfect tunneling can be changed into the perfect CAR. With the further increase of $\Gamma$,
the perfect CAR can keep very well with $T^{CAR}=1$ and $T=0$.

Let us study the effect of the position of the STM tip on the perfect CAR.
Fig.2 (c) shows the CAR coefficient $T^{CAR}$
versus the coupling position of the STM tip with several coupling sizes.
Here the coupling size of the STM tip $n\times n$ (with $n=3$, $4$ and $5$) means
that there are $n\times n =n^2$ sites in the TSC/QAHI ribbon coupled to the STM tip.
From Fig.2 (c), one can see that $T^{CAR}=1$ always regardless of
the coupling position and the size of the STM tip. That is that
the perfect CAR can always take place and is robust against
the coupling position and the size of the STM tip.
It's worth mentioning that the size of STM tip usually is very small in general
STM spectroscopy experiments.
However, the size of the present setup is in micrometer level.\cite{HeQL}
So the size of the STM tip is also required large,
e.g. about 100nm or larger. In usual, a large STM tip should
be easy to realize experimentally.

\begin{figure}[tbhp]
\centering
\includegraphics[width=.9\linewidth]{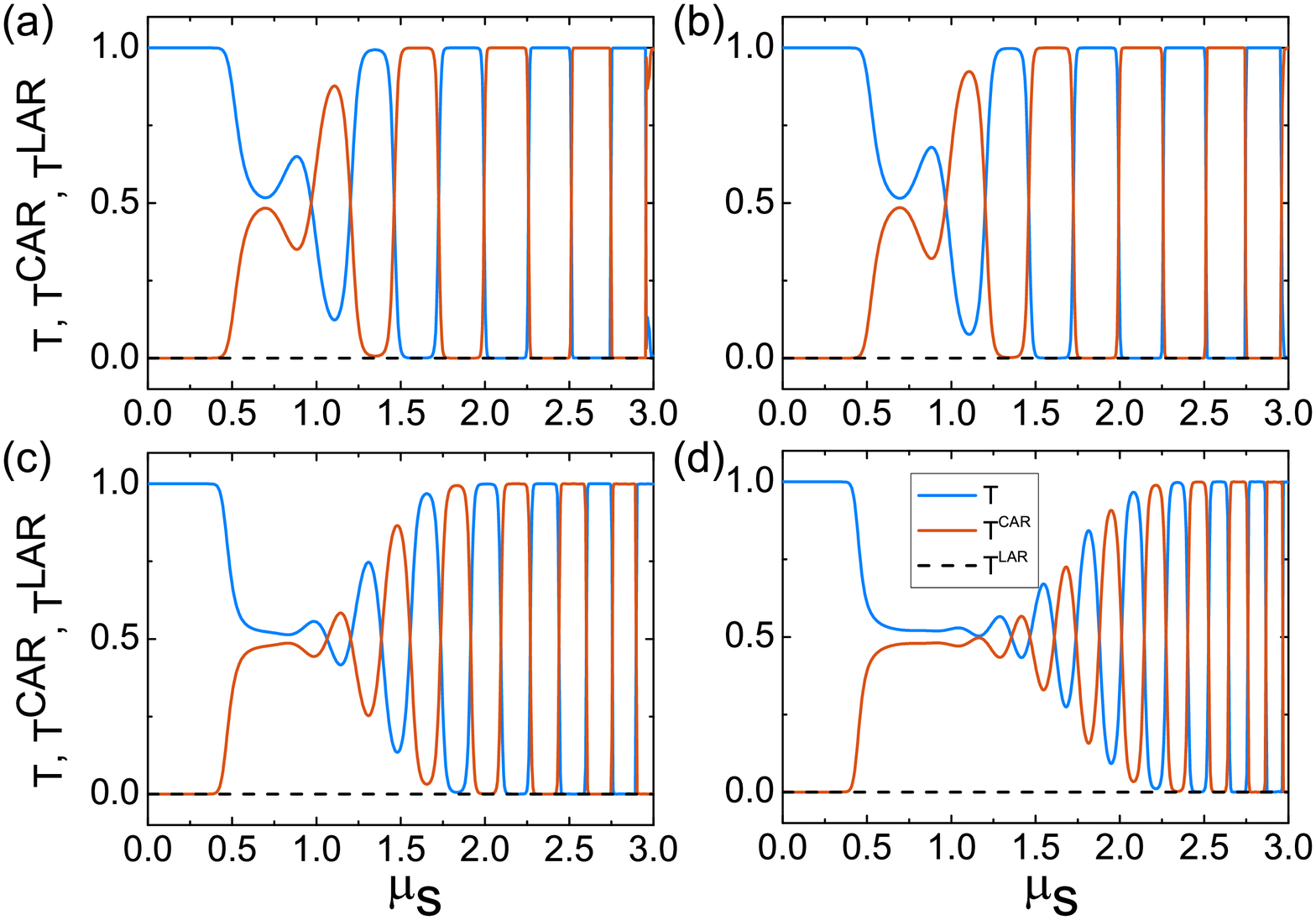}
\caption{(color online) $T$, $T^{CAR}$ and $T^{LAR}$ versus $\mu_s$
for the different size of the TSC island, $(L_x, L_y)= (20a, 80a)$ in (a), $(L_x, L_y)= (20a, 100a)$ in (b), $(L_x, L_y)= (30a, 90a)$ in (c), and $(L_x, L_y)= (40a, 90a)$ in (d), with
$L_x$ and $L_y$ being the length and width of the TSC island.
All the unmentioned parameters are the same as Fig.1 (c).
}
\label{FigS1}
\end{figure}

\begin{figure}[tbhp]
\centering
\includegraphics[width=.75\linewidth]{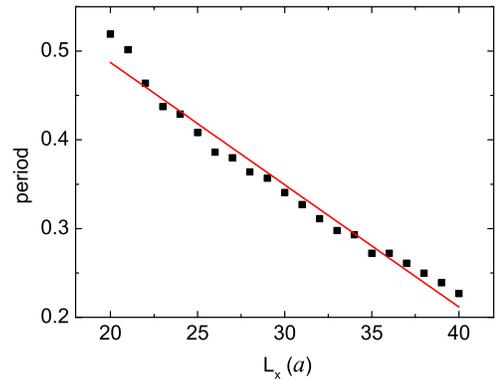}
\caption{(color online)
The switching period of the $T^{CAR}$ with respect to $\mu_s$ as a function of the length
$L_x$ of the TSC island.
The black solid squares are extracted from the curves of the CAR coefficient versus
$\mu_s$ and the red solid line is the linear regression for the discrete black solid squares.
All the unmentioned parameters are the same as Fig.1(c). }
\label{FigS2}
\end{figure}

Let us study the effect of the size of the TSC island on the regulation of
the phase of Majorana state and quantized perfect CAR.
Fig.3 shows $T$, $T^{CAR}$ and $T^{LAR}$ as functions of $\mu_s$ for different sizes of
the TSC island.
As the length $L_x$ and width $L_y$ of the TSC island change,
the $T=1$ plateaus and $T^{CAR}=1$ plateaus can well remain, and they still appear alternately.
That is that the perfect tunneling and the perfect CAR effect can occur regardless of
the size of the TSC island.
The longer the length $L_x$ of the TSC island is, the more frequent the alternation between
the perfect tunneling and perfect CAR is.
Moreover, Fig.4 shows the switching period of $T^{CAR}$
with respect to $\mu_s$ as a function of the length $L_x$ of the TSC island.
It can be seen that the period is linearly determined by $L_x$.
On the other hand, without the coupling of the STM tip, $T=1$ and $T^{CAR}=T^{LAR}=0$ always as shown in Fig.1 (b),
no matter what $\mu_s$ and TSC island size are.
This indicates that the coupling of the STM tip can
well regulate the phase of the Majorana state from $0$ to $\pi$,
which is independent of the size of the TSC island.

\begin{figure*}
\centering
\includegraphics[width=15cm]{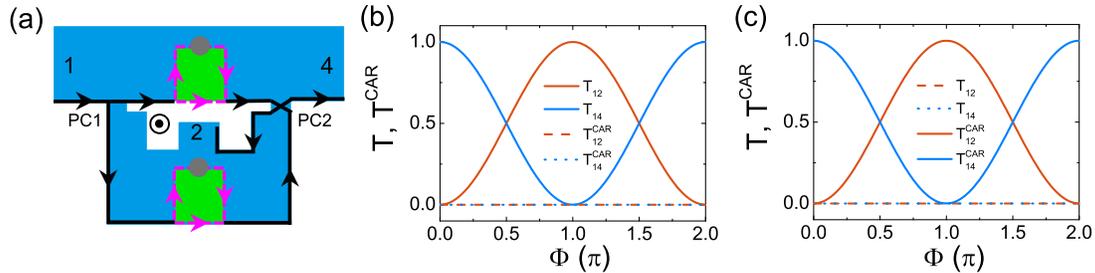}
\caption{(color online) (a) The configuration of the electronic Mach-Zehnder interferometer.
PC1 and PC2 are two point contacts connecting the two TSC/QAHI junctions
and a magnetic flux $\Phi$ is threaded in the cavity.
The widthes of terminals 1, 2 and 4 are $150a$, $70a$ and $150a$,
and the size of the lower TSC/QAHI junction is $(20a, 90a)$.
(b) and (c) show $T_{1n}$ and $T^{CAR}_{1n}$ from terminal 1 to terminal $n$ ($n=2,4$)
as functions of $\Phi$ with $\mu_s= 1.87$ (b) and $2.1$ (c),
which correspond to the perfect tunneling and perfect CAR regimes.
All the other unmentioned parameters are the same as Fig.1 (c).
}
\label{fig:fig3}
\end{figure*}

In addition, we also study the effect of the TSC gap $\Delta$ and the chemical potential $\mu_{QAHI}$ of the QAHI region on the regulation of the phase of Majorana state and the perfect CAR.
The perfect tunneling and perfect CAR can always survive as long as $\mu_{QAHI}$
in the bulk gap of the QAHI region, and they can also hold
in a wide range of the TSC gap $\Delta$.
Hence, the perfect CAR should be easily observed in the experiment and it can be a solid proof
for the existence of CMEM.

\subsection{Coherence}

With the coupling of the STM tip, the electron and hole may go into and then come back from the STM tip
which is akin to the B\"{u}ttiker virtual probes. Can it cause the dephasing?
Next, we study whether the outgoing electrons or holes remain coherent
by using an electronic Mach-Zehnder interferometer with two point contacts (PC1 and PC2)
as shown in Fig.5 (a).\cite{Ji} In the PCs, by fine tuning the coupling strength,
an incident electron is equally transmitted to two paths, similar to the beam splitters.
Two TSC islands are introduced in the transmission pathes of the interferometer
and a magnetic flux $\Phi$ is threaded in the cavity. The PC1 splits the incoming edge current
from terminal 1 into two paths respectively. After crossing the TSC islands, they recombine
again in PC2, and finally go to the terminal 2 and 4. Fig.5(b) and 5(c) show the normal
tunneling coefficient $T_{1n}$ and CAR coefficient $T^{CAR}_{1n}$ ($n=2,4$)
from the terminal 1 to the terminal $n$ as functions of magnetic flux $\Phi$
which introduces a phase difference between the two paths via the Aharonov-Bohm effect.
In the parameter regimes of Fig.5(b) and 5(c), there are the occurrences of the perfect tunneling
and perfect CAR in the region of the TSC/QAHI junction respectively,
with the outgoing particles being electron and hole correspondingly. As shown in Fig.5 (b) [Fig.5 (c)],
the normal tunneling coefficient $T_{1n}$ (the CAR coefficient $T^{CAR}_{1n}$) oscillates between 0 and 1
with the increase of the magnetic flux $\Phi$, but $T_{12}+T_{14}=1$ and $T^{CAR}_{12} = T^{CAR}_{14} =0$
($T^{CAR}_{12}+T^{CAR}_{14}=1$ and $T_{12} = T_{14} =0$).
The oscillating amplitudes of $T_{1n}$ and $T^{CAR}_{1n}$ being 1 indicates that the electron
and hole scattered off the TSC island are still completely coherent despite of the coupling of the STM tip.
The survival of the phase coherence results from that the Majorana state propagating forward
along the CMEMs can only take a phase $0$ or $\pi$, and the phase cannot be changed randomly
by the coupling of the STM tip\cite{decoherence}.

\subsection{Experimental signature}

Finally, we study the physically observable quantities caused by the perfect CAR and
the adjustment of the phase of the Majorana state. We consider a six-terminal Hall bar as shown in Fig.6 (a).
Here the TSC and lead 4 are grounded, and a small bias $V$ is applied to lead 1
with $V_s=V_4=0$ and $V_1=V$. The other leads are the voltage probes with zero current.
Without coupling of the STM tip, the perfect normal tunnelings occur along the QAHI edge
and the CMEM with $T^{CAR}=T^{LAR}=0$ as shown in Fig.1 (b). In this case, the transport properties are
completely the same as the quantum anomalous Hall effect with $V_2=V_3=V_4=0$, $V_5=V_6=V_1=V$,
$I_1=-I_4 =(e^2/\hbar)V$, the longitudinal resistances $V_{23}/I_1 = V_{65}/I_1 =0$ and
the Hall resistances $V_{62}/I_1 = V_{53}/I_1 =\hbar/e^2$ regardless of the chemical potential $\mu_s$,
where $V_{nm}\equiv V_n -V_m$. However, with the coupling of the STM tip,
the results are essentially different, and the observable quantities mentioned above are strongly dependent on the chemical potential
$\mu_s$ as shown in Fig.6 (b-d). Now the perfect tunneling and perfect CAR occur alternately
when the TSC island is in the chiral TSC phase with $\mathcal{N}=1$. For the perfect tunneling,
the results are the same as above. But for the perfect CAR, the voltage of the lead 5 is $V_5=-V$
instead of $V$ [see Fig.6 (b)] because the CAR coefficient $T^{CAR}=1$.
Notice that $V_5$ is negative and
lower than the voltages of the leads 1 and 4, then a pumping current will be driven.
The current $I_4=I_1 =(e^2/\hbar)V$ as shown in Fig.6 (c), in which the currents at both leads 1 and 4
flow into the center region.
As there is no external power in the right circuit loop consisting of the lead 4 and superconductor lead in Fig.6 (a),
this is a pumping current, where the electrons in the lead 1 draw those in the lead 4 to combine into Cooper pairs
and enter into the superconductor lead eventually. The longitudinal resistance of the upper edge $V_{23}/I_1 =0$
which is independent of $\mu_s$, because of unidirectionality of the QAHI edge states.
Nevertheless, the longitudinal conductance of the lower edge, $I_1/V_{65} =\frac{1}{2}e^2/h$
at the half-integer quantized value [Fig.6 (d)], due to the occurrence of the perfect CAR.
The left Hall resistance $V_{62}/I_1 =h/e^2$, but the right Hall resistance $V_{53}/I_1 = h/e^2$
for the perfect tunneling and $-h/e^2$ for the perfect CAR [see Fig.6 (d)].
These results can give a solid proof for the CMEM. Moreover, considering that
the proposed setup is very similar to the one in the recent experiment\cite{HeQL},
the predicted perfect CAR should be experimentally observed in the present technologies.

\begin{figure}
\centering
\includegraphics[width=.9\linewidth]{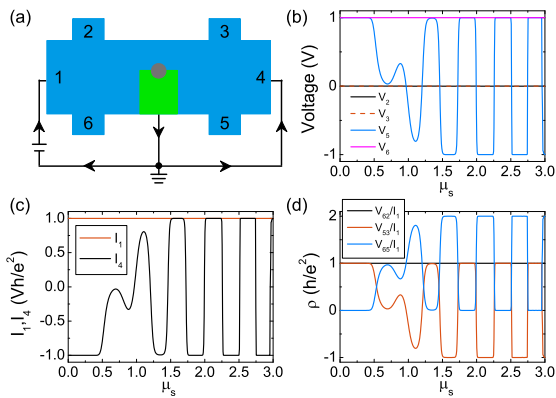}
\caption{(color online) (a) the schematic diagram of a six-terminal Hall bar system consisting of the QAHI and TSC.
(b-d) show the voltages $V_n$ (b), the currents $I_1$ and $I_4$ (c), and the longitudinal and Hall resistances (d)
as functions of $\mu_s$.
The widthes of leads 2, 3, 5, and 6 are $70a$, and all the unmentioned parameters are the same as Fig. 1 (c).
}
\label{fig:fig4}
\end{figure}

In a realistic experiment involving the STM tip, the voltage and conductance of the STM tip
would be there for direct readout. Whether there exist signatures of the phase adjustment
from these experimental measurements.
We calculate both the voltage $V_{tip}$ of the STM tip with its current being zero
and the conductance $dI_{tip}/dV$ at the zero voltage $V_{tip}=0$.
The results show that the voltage $V_{tip}$
and the conductance $d I_{tip}/dV$ are zero for both $T=1$ and $T^{CAR}=1$ phases, but non-zero $V_{tip}$ and $d I_{tip}/dV$ appear
at the transitions between $T=1$ and $T^{CAR}=1$ (see the Appendix B).
This means that the measurements from the STM tip can only manifest the
transition between the two phases, but cannot distinguish them.

\section{\label{sec4}conclusion}

In summary, we have studied the effect of a STM tip on the chiral Majorana edge modes
and demonstrated that the phase of the Majorana states can be regulated by the coupling
of the STM tip. When a phase $\pi$ is introduced for Majorana state,
a perfect CAR occurs and all other scattering processes completely disappear.
Moreover, the outgoing electrons and holes can keep the phase coherence well despite of the STM tip.
The physically observable consequences from the perfect CAR are also studied in a six-terminal Hall bar setup,
in which the longitudinal and Hall resistances show the quantized plateaus. These findings can give
an undoubted proof for the existence of the chiral Majorana edge mode
and open up an avenue to control the phase of Majorana state.

\acknowledgments
This work was financially supported by National Key R and D Program of China (2017YFA0303301),
NBRP of China (2015CB921102), NSF-China under Grants Nos. 11474084 and 11574007,
and the Key Research Program of the Chinese Academy of Sciences (Grant No. XDPB08-4).

\begin{appendix}
\section{Derivation of transport formula}

By using the nonequilibrium Green's function method,
the normal tunneling, CAR and LAR coefficients can be obtained from:\cite{asun2,asun3}
\begin{align} \label{eqn:S2}
\begin{split}
   \tilde{T}_{nm}(E) & =\mathrm{Tr}[{\bf \Gamma}_{ee}^n {\bf G}_{ee}^r
   {\bf \Gamma}_{ee}^m {\bf G}_{ee}^a], \\
    \tilde{T}_{nm}^A(E) & =\mathrm{Tr}[{\bf \Gamma}_{ee}^n {\bf G}_{eh}^r
    {\bf \Gamma}_{hh}^m {\bf G}_{he}^a],
\end{split}
\end{align}
where ``e/h" represent electron/hole respectively, $E$ is the incident
energy, $n$ and $m$ are the index of the terminals, including
the left and right terminals in Fig.1 (a), the terminal $n$ ($n=1,2,...,6$) in Fig.6 (a), and the STM tip terminal.
${\bf G}^r(E)=[E-\mathcal{H}_{\mathrm{BdG}}-\sum_{n} {\bf \Sigma}^r_{n}]^{-1}$
is the retarded Green's function,
where $\mathcal{H}_{\mathrm{BdG}}$ is the BdG Hamiltonian of the central region.
$ {\bf \Gamma}^{n}(E)=i[ {\bf \Sigma}_{n}^r- {\bf \Sigma}_{n}^a]$ is the line-width function.
The self-energy ${\bf \Sigma}_{n}^r={\bf \Sigma}_{n}^{a\dagger}$ stems from the coupling between the
terminal $n$ and the center regions.
For the QAHI terminal, the self-energy can be calculated numerically\cite{alee}.
While for the STM tip terminal, ${\bf \Sigma}_{tip}^r = -\frac{i\Gamma}{2}\mathbf{I}_{4N}$,
where $\Gamma=2\pi\rho t_d^2$ is the coupling strength between the STM tip and the TSC island,
and $\mathbf{I}_{4N}$ is the $4N \times 4N$ unit matrix in the BdG representation and
$N$ is the number of the sites coupled with the STM tip.
In Eq.\ref{eqn:S2}, $\tilde{T}_{nm}(E)$ ($n\not= m$) is the normal tunneling coefficient from
the terminal $n$ to the terminal $m$, and $\tilde{T}^A_{nm}(E)$ is the Andreev reflection coefficient.
For $n\not= m$, $\tilde{T}^A_{nm}$ is the CAR coefficient, while for $n=m$ it is the LAR coefficient.
Due to there being only one edge mode in the QAHI terminal, the normal reflection coefficient for the
QAHI terminal is $\tilde{R}_{nn} =1 -\sum\limits_{m(m\not= n)}\tilde{T}_{nm} -\sum\limits_{m}\tilde{T}^A_{nm}$.

After obtaining these transmission coefficients, the current in the terminal $n$ at the small bias limits
can be calculated from multi-probe Landauer-B\"{u}ttiker formula\cite{asun1},
\begin{align} \label{eqn:S3}
\begin{split}
   I_n= & (e^2/h)\left[(V_n-V_s)\tilde{T}_{sn}(0)+ \sum_{m(m\not= n)}(V_n-V_m)\tilde{T}_{mn}(0)
   \right. \\
     & \left. + 2V_n\tilde{T}^A_{nn}(0)  + \sum_{m(m\not= n)}(V_n+V_m)\tilde{T}^A_{mn}(0) \right],
\end{split}
\end{align}
where $V_n$ is the voltage of terminal $n$.
Here the voltage $V_s$ of the superconductor lead is set to zero.
While the incident energy $E=0$ which is inside the superconductor gap,
the tunneling coefficient $\tilde{T}_{sn}$ between terminal $n$ and superconductor lead is zero,
so the first term in Eq.(\ref{eqn:S3}) vanishes.
For the voltage terminals [e.g. the STM tip, the terminal 2, 3, 5 and 6 in Fig.6 (a)],
the currents are zero, then their voltage can be solved from the Eq.(\ref{eqn:S3}).

We take the system in Fig.1 (a) as an example.
Setting the voltages $V_L$ and $V_R$ for left and right QAHI terminals, and $I_{tip}=0$ in the STM tip,
the currents $I_L$, $I_R$ and voltage $V_{tip}$ can be easily solved from Eq.(\ref{eqn:S3}),
and they are linearly dependent on $V_L$ and $V_R$. For example, the current of the right terminal
can be written as
$I_R=(e^2/h)(a_R V_R+a_L V_L)=(e^2/h)[\frac{a_R+a_L}{2}(V_R+V_L)+\frac{a_R-a_L}{2}(V_R-V_L)]$.
Here the coefficients $\frac{a_R-a_L}{2}$ ($\frac{a_R+a_L}{2}$) of $V_R-V_L$ ($V_R+V_L$) represent
the probability of the outgoing electron (hole), which is the total normal tunneling coefficient $T$
(total CAR coefficient $T^{CAR}$), including the direct tunneling from $L$ to $R$,
the indirect process from $L$ passing the STM tip to $R$, and so on.
Besides, the LAR coefficient $T^{LAR} =\tilde{T}^{A}_{RR}$ and the normal reflection $R=1-T-T^{CAR}-T^{LAR}$.
Both of them are zero because of no touch between the TSC island and the upper edge of the QAHI ribbon [see Fig.1 (a)].

\section{Voltage and conductance of the STM tip}

Let us study both the voltage $V_{tip}$ of the STM tip while its current being zero
and the conductance $dI_{tip}/dV$ at the zero voltage $V_{tip}=0$.
Here the setup as shown in Fig.1(a) is considered.
The voltage of the left QAHI terminal sets $V$, and the voltages of the TSC island and
the right QAHI terminal are zero.
For the STM tip terminal, two boundary conditions are considered:
1) The current of the STM tip sets to zero (i.e. the open circuit condition),
the voltage $V_{tip}$ of the STM tip is studied.
2) The voltage $V_{tip}$ is zero (i.e. ground), the conductance $dI_{tip}/dV$
is investigated.

\begin{figure}[htpb]
\centering
\includegraphics[width=.9\linewidth]{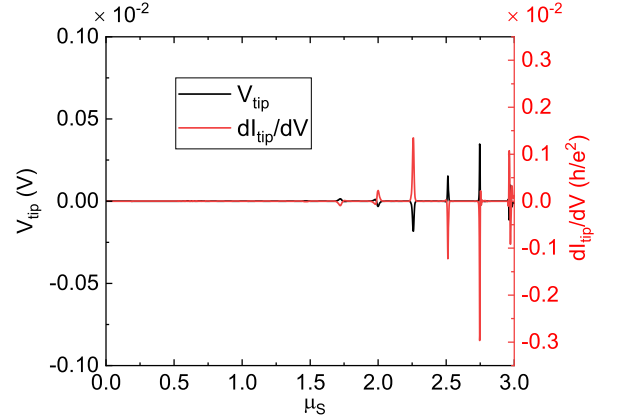}
\caption{ (color online)  The voltage of the STM tip $V_{tip}$ and the conductance $d I_{tip}/dV$ as a function of $\mu_s$.
The voltage $V_{tip}$ is calculated at the open circuit condition with the STM tip current $I_{tip}=0$.
The conductance $d I_{tip}/dV$ is calculated at $V_{tip}=0$.
The voltages $V_L = V$ and $V_R = V_s=0$.
The parameters are the same as Fig.1(c).}
\end{figure}

Fig.7 shows the voltage $V_{tip}$ and the conductance $dI_{tip}/dV$ as
a function of the chemical potential $\mu_s$.
Here the parameters are the same as Fig.1(c).
The voltage $V_{tip}$ and the conductance $d I_{tip}/dV$ are zero for both
$T=1$ and $T^{CAR}=1$ phases. But non-zero $V_{tip}$ and $d I_{tip}/dV$ appear
at the transitions between $T=1$ and $T^{CAR}=1$.
These results are well consistent with
the physical picture of the one-dimensional CMEMs as shown in Fig.1(a).
From the one-dimensional CMEMs in the Fig.1(a), one can see that only the Majorana fermion
$\gamma_2$ [$\gamma_2=\frac{1}{2i}(\eta_L-\eta_L^{\dagger}$)] can tunnel into the STM tip.
So the normal tunneling and Andreev reflection coefficients are
$\tilde{T}_{L,tip} = \tilde{T}^A_{L,tip} =|c|^2/4$ and $\tilde{T}_{R,tip} = \tilde{T}^A_{R,tip} =0$,
where $c$ is the tunneling amplitude from $\gamma_2$ to the STM tip.
$c$ is positive or negative for the $T=1$ or $T^{CAR}=1$ phase.
Then from multi-probe Landauer-B\"{u}ttiker formula, we have:
\begin{align} \label{eqn:S4}
\begin{split}
  I_{tip} = & (e^2/h)
   \left[(V_{tip}-V_L)\tilde{T}_{L,tip}+2V_{tip}\tilde{T}^A_{tip,tip} \right.\\
    & \left.+ (V_{tip}+V_L)\tilde{T}^A_{L,tip}\right] \\
    =& 2(e^2/h) V_{tip} \left( \tilde{T}^A_{tip,tip} + |c|^2/4\right).
 \end{split}
\end{align}
So the tip voltage $V_{tip}$ is zero at the open circuit condition with the current $I_{tip}=0$,
and the current $I_{tip}=0$ (i.e. the conductance $d I_{tip}/dV =0$) while the tip sets ground with $V_{tip}=0$.

\end{appendix}


\begin{thebibliography}{}
\bibitem{ReadN}
N. Read and D. Green, Phys. Rev. B {\bf 61}, 10267 (2000).

\bibitem{QiXL1}
X.-L. Qi and S.-C. Zhang, Rev. Mod. Phys. {\bf 83}, 1057 (2011).

\bibitem{Alicea}
J. Alicea, Rep. Prog. Phys. {\bf 75}, 076501 (2012).

\bibitem{Beenakker}
C.W.J. Beenakker, Annu. Rev. Condens. Matter Phys. {\bf 4}, 113-136 (2013).

\bibitem{ElliottSR}
S.R. Elliott and  M. Franz,  Rev. Mod. Phys. {\bf 87}, 137 (2015).

\bibitem{IvanovDA}
D. A. Ivanov, Phys. Rev. Lett. {\bf 86}, 268 (2001).

\bibitem{NayakC}
C. Nayak, S.H. Simon, A. Stern, M. Freedman, and S. Das Sarma, Rev. Mod. Phys. {\bf 80}, 1083 (2008)

\bibitem{FuL2}
L. Fu and C.L. Kane, Phys. Rev. B {\bf 79}, 161408 (2009).

\bibitem{Lutchyn}
R.M. Lutchyn, J.D. Sau, and S. Das Sarma, Phys. Rev. Lett. {\bf 105}, 077001 (2010).

\bibitem{Nilsson}
J. Nilsson, A.R. Akhmerov, and C.W.J. Beenakker, Phys. Rev. Lett. {\bf 101}, 120403 (2008).

\bibitem{Law}
K.T. Law, P.A. Lee, and T.K. Ng, Phys. Rev. Lett. {\bf 103}, 237001 (2009).

\bibitem{WangZ}
Z. Wang, X.-Y. Hu, Q.-F. Liang, and X. Hu, Phys. Rev. B {\bf 87}, 214513 (2013).

\bibitem{QiXL2}
X.-L. Qi, T.L. Hughes, and S.-C. Zhang, Phys. Rev. B {\bf 82}, 184516 (2010).

\bibitem{WangJ}
J. Wang, Q. Zhou, B. Lian, and S.-C. Zhang, Phys. Rev. B {\bf 92}, 064520 (2015).

\bibitem{FuL1}
L. Fu and C.L. Kane, Phys. Rev. Lett. {\bf 100}, 096407 (2008).

\bibitem{SauJD}
J.D. Sau, R.M. Lutchyn, S. Tewari, and S. Das Sarma, Phys. Rev. Lett. {\bf 104}, 040502 (2010).

\bibitem{Chung}
S.B. Chung, X.-L. Qi, J. Maciejko, and S.-C. Zhang, Phys. Rev. B {\bf 83}, 100512 (2011).

\bibitem{Lian}
B. Lian, J. Wang, and S.-C. Zhang, Phys. Rev. B {\bf 93}, 161401 (2016).

\bibitem{nadd6}
J. Wang, Phys. Rev. B {\bf 94}, 214502 (2016).

\bibitem{HeQL}
Q.L. He, L. Pan, A.L. Stern, E.C. Burks, X. Che, G. Yin, J. Wang, B. Lian, Q. Zhou, E.S. Choi, K. Murata,
X. Kou, Z. Chen, T. Nie, Q. Shao, Y. Fan, S.-C. Zhang, K. Liu, J. Xia, and K.L. Wang, Science {\bf 357}, 294 (2017).

\bibitem{nadd7}
C.-Z. Chen, J. J. He, D.-H. Xu, and K. T. Law,
Phys. Rev. B {\bf 96}, 041118 (2017).


\bibitem{GMenard}
G.C. M\'{e}nard, S. Guissart, Ch. Brun, R.T. Leriche, M. Trif,
F. Debontridder, D. Demaille, D. Roditchev, P. Simon, and T. Cren, Nature Commun. {\bf 8}, 2040 (2017)

\bibitem{Tiwari}
R.P. Tiwari, U. Z$\ddot{u}$licke, and C. Bruder, New J. Phys. {\bf 16}, 025004 (2014).

\bibitem{Andreev}
A.F. Andreev, Sov. Phys. JETP {\bf 19}, 1228 (1964).

\bibitem{asun1}
Q.-F. Sun, J. Wang, and T.-H Lin, Phys. Rev. B {\bf 59}, 3831 (1999).

\bibitem{Byers}
J.M. Byers and M.E. Flatt\'{e}, Phys. Rev. Lett. {\bf 74}, 306 (1995).

\bibitem{Deutscher}
G. Deutscher and D. Feinberg, Appl. Phys. Lett. {\bf 76}, 487 (2000).

\bibitem{nadd2}
Z. Hou, Y. Xing, A.-M. Guo, and Q.-F. Sun, Phys. Rev. B {\bf 94}, 064516 (2016);
Y.-T. Zhang, X. Deng, Q.-F. Sun, and Z. Qiao, Sci. Rep. {\bf 5}, 14892 (2015).

\bibitem{Recher}
P. Recher, E.V. Sukhorukov, and D. Loss, Phys. Rev. B {\bf 63}, 165314 (2001).

\bibitem{nadd1}
Q.-F. Sun, H. Guo, and T.-H. Lin, Phys. Rev. Lett. {\bf 87}, 176601 (2001).

\bibitem{Chen}
W. Chen, R. Shen, L. Sheng, B.G. Wang, and D.Y. Xing, Phys. Rev. B {\bf 84}, 115420 (2011).

\bibitem{Veldhorst}
M. Veldhorst and A. Brinkman, Phys. Rev. Lett. {\bf 105}, 107002 (2010).

\bibitem{Cayssol}
J. Cayssol, Phys. Rev. Lett. {\bf 100}, 147001 (2008).

\bibitem{James}
J.J. He, J. Wu, T.-P. Choy, X.-J. Liu, Y. Tanaka, and K. T. Law,  Nat. Commun. {\bf 5}, 3232 (2014).

\bibitem{ZhangYT}
Y.T. Zhang, Z. Hou, X.C. Xie, and Q.-F. Sun, Phys. Rev. B {\bf 95}, 245433 (2017).

\bibitem{WangJ2}
J. Wang, L. Hao, and K.S. Chan, Phys. Rev. B {\bf 91}, 085415 (2015).

\bibitem{addDLoss}
C. Schrade, A.A. Zyuzin, J. Klinovaja, and D. Loss, Phys. Rev. Lett. {\bf 115}, 237001 (2015);
C. Reeg, J. Klinovaja, and D. Loss, Phys. Rev. B {\bf 96}, 081301 (2017).

\bibitem{addPachos}
C. Benjamin and J.K. Pachos, Phys. Rev. B {\bf 81}, 085101 (2010).

\bibitem{nadd3}
Q.-F. Sun, Y.-X. Li, W. Long, and J. Wang, Phys. Rev. B {\bf 83}, 115315 (2011).

\bibitem{nadd4}
Z. Hou and Q.-F. Sun, Phys. Rev. B {\bf 96}, 155305 (2017).

\bibitem{nadd5}
K. Zhang, J. Zeng, Y. Ren, and Z. Qiao, Phys. Rev. B {\bf 96}, 085117 (2017).

\bibitem{FuL3}
L. Fu and C.L. Kane, Phys. Rev. Lett. {\bf 102}, 216403 (2009).

\bibitem{Akhmerov}
A.R. Akhmerov, J. Nilsson, and C.W.J. Beenakker, Phys. Rev. Lett. {\bf 102}, 216404 (2009).

\bibitem{Park}
S. Park, J.E. Moore, and H.-S. Sim, Phys. Rev. B {\bf 89}, 161408 (2014).

\bibitem{nadd8}
H.-H. Sun, K.-W. Zhang, L.-H. Hu, C. Li, G.-Y. Wang,
H.-Y. Ma, Z.-A. Xu, C.-L. Gao, D.-D. Guan, Y.-Y. Li, C. Liu, D. Qian,
Y. Zhou, L. Fu, S.-C. Li, F.-C. Zhang, and J.-F. Jia, Phys. Rev.
Lett. {\bf 116}, 257003 (2016);
L.-H. Hu, C. Li, D.-H. Xu, Y. Zhou, and F.-C. Zhang, Phys. Rev. B {\bf 94}, 224501 (2016).

\bibitem{nadd9}
P. Devillard, D. Chevallier, and M. Albert, Phys. Rev. B {\bf 96}, 115413 (2017).

\bibitem{ChangCZ1}
C.-Z. Chang, J. Zhang, X. Feng, J. Shen, Z. Zhang, M. Guo,
K. Li, Y. Ou, P. Wei, L.-L. Wang, Z.-Q. Ji, Y. Feng, S. Ji,
X. Chen, J. Jia, X. Dai, Z. Fang, S.-C. Zhang, K. He, Y. Wang, L. Lu,
X.-C. Ma, Q.-K. Xue, Science {\bf 340}, 167 (2013).

\bibitem{Checkelsky}
J.G. Checkelsky, R. Yoshimi, A. Tsukazaki, K.S. Takahashi, Y. Kozuka, J. Falson, M. Kawasaki,
and Y. Tokura, Nat. Phys. {\bf 10}, 731 (2014).

\bibitem{Kou}
X. Kou, S.-T. Guo, Y. Fan, L. Pan, M. Lang, Y. Jiang, Q. Shao, T. Nie,
K. Murata, J. Tang, Y. Wang, L. He, T.-K. Lee, W.-L. Lee, and K.L. Wang, Phys. Rev. Lett. {\bf 113}, 137201 (2014).

\bibitem{Bestwick}
A.J. Bestwick, E.J. Fox, X. Kou, L. Pan, K.L. Wang, and D. Goldhaber-Gordon, Phys. Rev. Lett. {\bf 114}, 187201 (2015).

\bibitem{Kandala}
A. Kandala, A. Richardella, S. Kempinger, C.-X. Liu, and N. Samarth, Nature Commun. {\bf 6}, 7434 (2015).

\bibitem{ChangCZ2}
C.-Z. Chang, W. Zhao, D.Y. Kim, H. Zhang, B.A. Assaf, D. Heiman,
S.-C. Zhang, C. Liu, M.H.W. Chan, and J.S. Moodera, Nat. Mater. {\bf 14}, 473 (2015).

\bibitem{adatta}
S. Datta, {\sl Electronic transport in mesoscopic system},
Cambridge University Press, Bambridge, 1995.

\bibitem{Schnyder}
A.P. Schnyder, S. Ryu, A. Furusaki, and A.W.W. Ludwig, Phys. Rev. B {\bf 78}, 195125 (2008)

\bibitem{asun2}
Q.-F. Sun and X.C. Xie, J. Phys.: Condens. Matter {\bf 21}, 344204 (2009).

\bibitem{asun3}
S.-G. Cheng, Y. Xing, J. Wang, and Q.-F. Sun, Phys. Rev. Lett. {\bf 103}, 167003  (2009).

\bibitem{alee}
D.H. Lee and J.D. Joannopoulos, Phys. Rev. B {\bf 23}, 4997 (1981).

\bibitem{Ji}
Y. Ji, Y. Chung, D. Sprinzak, M. Heiblum, D. Mahalu, and H. Shtrikman, Nature {\bf 422}, 415 (2003).

\bibitem{decoherence}
There inevitably exist some other potential sources of decoherence in real experiments.
As the Majorana fermions are charge neutral particles, the coupling of the Majorana fermion
with the surrounding environment would be weak and the dephasing from them is neglectable.
Moreover, the dephasing effect of the interaction within the Majorana fermions will be suppressed
to a great extent at the low temperature\cite{FuL3}. Considering these reasons, the effects of
these dephasing processes are not considered here.

\end{thebibliography}
\end{document}